\documentclass{mn2e}
\usepackage{natbib}
\usepackage{graphicx}
\usepackage{pslatex}

\begin{document}


\title{Luminosity function, sizes and FR dichotomy of radio-loud AGN}

\author[C.R. Kaiser \& P.N. Best]{Christian R. Kaiser$^1$\thanks{crk@soton.ac.uk} and Philip N. Best$^2$\\
$^1$ School of Physics \& Astronomy, University of Southampton, Southampton SO17 1BJ\\
$^2$ Institute for Astronomy, Royal Observatory Edinburgh, Blackford Hill, Edinburgh EH9 3HJ
}

\maketitle

\begin{abstract}
The radio luminosity function (RLF) of radio galaxies and radio-loud quasars is often modelled as a broken power-law. The break luminosity is close to the dividing line between the two Fanaroff-Riley (FR) morphological classes for the large-scale radio structure of these objects. We use an analytical model for the luminosity and size evolution of FRII-type objects together with a simple prescription for FRI-type sources to construct the RLF. We postulate that all sources start out with an FRII-type morphology. Weaker jets subsequently disrupt within the quasi-constant density cores of their host galaxies and develop turbulent lobes of type FRI. With this model we recover the slopes of the power laws and the break luminosity of the RLF determined from observations. The rate at which AGN with jets of jet power $Q$ appear in the universe is found to be proportional to $Q^{-1.6}$. The model also roughly predicts the distribution of the radio lobe sizes for FRII-type objects, if the radio luminosity of the turbulent jets drops significantly at the point of disruption. We show that our model is consistent with recent ideas of two distinct accretion modes in jet-producing AGN, if radiative efficiency of the accretion process is correlated with jet power. 
\end{abstract}


\section{Introduction}

The large scale structure, or lobes, produced by the jets of radio
galaxies and radio-loud quasars shows a large range of different
morphologies \citep[for an overview see][]{mg91}. Nevertheless they can be
grouped into two main classes according to their surface brightness
structure \citep{fr74}. Objects in Fanaroff-Riley class I, FRI for short,
are brightest at their centres while FRII-type objects are
edge-brightened. The two FR classes are also broadly separated
in terms of their radio luminosity. The original dividing line was placed
at $5\times 10^{25}$\,W\,Hz$^{-1}$ at an observing frequency of 178\,MHz,
but was later shown to depend on the properties of the host galaxies
\citep{lo96}.

Higher angular resolution observations revealed that the more luminous
FRII-type sources contain laminar jets extending from the AGN to very
bright radio hotspots enclosed in diffuse, low surface brightness
lobes. The morphology is consistent with the interpretation of the
hotspots as the impact sites of the jets on the ambient gas and the lobes
as the `waste baskets' for the material and energy transported by the jets
\citep{ps74}. The FRI class is made up of sources with different
morphologies. However, the majority shows jets that appear laminar in the
innermost region inflating turbulent lobes after passing through a flare
point. A minority of FRI sources shows a `fat double' morphology which is
reminiscent of the diffuse lobes of the FRII class, but without
hotspots \citep{ol89,ow91}. In this paper we concentrate on the turbulent
FRI-types.

The luminosity dividing the FR classes is remarkably close to the break in
the radio luminosity function (RLF) \citep[e.g.][]{dp90}. While this has
been noted many times, most work on the RLF has focussed on its
cosmological evolution and its separation, not into FR classes, but into
contributions from starburst galaxies, flat and steep spectrum objects
\citep{jw99}. Both FR classes are part of the steep spectrum population
which also forms the parent population for the relativistically beamed
flat spectrum objects (flat-spectrum quasars and BL Lacs). Advances
in the modelling of the dynamics \citep[e.g.][hereafter KA]{sf91,ka96b}
and the synchrotron emission from the radio lobes of individual FRII-type
sources (\citealt{kda97a}, hereafter KDA; \citealt{brw99}; \citealt{mk02})
now allow us to investigate the shape of the RLF above the break
luminosity in detail. In this paper we present such an analysis. We argue
that FRI-type sources evolve out of FRII-type objects with weak jets. This
allows us to construct a self-consistent explanation for how the shape of
the RLF arises from the combination of both FR classes and the luminosity
evolution of individual sources.

Differences in the properties of radio sources were also shown to be
present beyond the radio waveband \citep[e.g.][]{hl79,zb95,lo96,ccc00,mhs04,hec06}. In particular the
difference in the emission properties of the AGN itself have let to the
idea of different accretion modes in low and high luminosity AGN
\citep{fkm04,hec07}. It is important to note that there is no one-to-one
correspondence between the emission properties and postulated accretion
mode of the AGN itself and the FR radio morphology. However, we show
that our model developed here extends the idea of different accretion
modes to include the FR dichotomy, if AGN with higher nuclear luminosities
also give rise to more powerful jets.

The observed distribution of the sizes of radio lobes of FRII-type objects
is difficult to reconcile with evolutionary models. There appear to be too
many small sources with sizes of tens of pc to a few kpc that cannot all
evolve into objects with lobes tens and hundreds of kpc across
\citep{ob97}. We show that our model can explain these observations, since
many small sources starting out with an FRII-type morphology `drop out' of
the observed samples by developing turbulent jets before their lobes grow
to large sizes.

In section \ref{evol} we review the luminosity and size evolution of
FRII-type objects. We present a simple model for the transition to and
subsequent evolution of FRI-type objects in section \ref{modfri}. Section
\ref{rlf} contains the construction of the RLF using these combined
models. The distribution of lobe sizes for the FRII-type objects is
discussed in section \ref{sizeII}. In section \ref{con} we fit our model
into the observational constraints from wavebands other than the radio and
discuss the proposal of two distinct accretion modes for jet-producing
AGN. Section \ref{sum} provides a summary of our main findings. Throughout
we use the definition of the spectral index $\alpha$ of synchrotron
emission as $F_{\nu} \propto \nu^{-\alpha}$.

\section{Evolutionary phases of a FRII-type radio-loud AGN}
\label{evol}

In this section we investigate the evolution of the radio luminosity of radio-loud AGN of type FRII as a function of source age and lobe size. This evolution is governed by the energy loss processes of the radiating relativistic electrons in the radio lobes. We are using the models of KA and KDA for this analysis and the relevant mathematical expressions are derived in the appendix. 

The key relation is given in equation (\ref{lumin}). Unfortunately, a general analytic solution of the integral in this equation is not possible. However, we can simplify the analysis considerably by setting the exponent of the initial power-law energy distribution of the relativistic electrons, $m$, to  2. This choice implies that the spectral index of the radio emission is fixed to $\alpha = \left( m -1 \right) /2 = 0.5$ in the case of negligible radiative energy losses of the relativistic electrons. We will show below that the spectral index also cannot become steeper than $\alpha = 1$ while the lobes remain observable in this idealised model. While these restrictions clearly limit the applicability of the model, the reduction in mathematical complexity allows us some important insights in the luminosity evolution.

Setting $m=2$ we solve the integral in equation (\ref{lumin}),
\begin{equation}
L_{\nu} = \frac{f_L}{1+\epsilon} Q p^{3/4} t \left( 1 - x_{\rm min}^{1+\epsilon} \right).
\end{equation}
We further simplify the analysis by assuming that the equation of state of the lobe material is dominated by the relativistic electrons and the magnetic field, i.e. $k=0$ and $\Gamma _l = 4/3$. Hence the exponent $\epsilon = a_1 /3$ (see equation \ref{eps}).

The value of $x_{\rm min}$ determines the fraction of the lobe volume that contributes to the emission at the observing frequency $\nu$. As we show in the appendix \ref{loss}, $x_{\rm min}$ is governed by the energy losses of the relativistic electrons in the lobe. The energy losses are due to the adiabatic expansion of the lobe, the emission of radio synchrotron radiation and the inverse Compton scattering of CMB photons. It is not possible to find an explicit expression for $x_{\rm min}$ taking into account all loss processes. However, while adiabatic losses affect the electrons at all times, the radiative loss processes usually dominate at different times during the source evolution. Hence we can define different phases during the evolution of a radio-loud AGN of type FRII during which different analytical expressions for their radio luminosity can be derived.

\subsection{Adiabatic losses}

We shall see below that energy losses due to the emission of synchrotron radiation affect the source evolution mainly at early times. Inverse Compton losses dominate at late times. Therefore in between these two regimes a source can go through a phase where only adiabatic losses are important. This is the simplest possibility and we therefore consider it first.

Neglecting radiative energy losses amounts to setting $g(x,t) =0$. Equation (\ref{xmin}) can then be solved to give 
\begin{equation}
x_{\rm min} = \left( \frac{\gamma_{\nu}}{\gamma _{\rm max}} \right)^{3 / a_1}.
\end{equation}
Unless the observing frequency is very high, we expect $\gamma_{\nu} \ll \gamma _{\rm max}$ and so the luminosity of the source in this phase is 
\begin{equation}
L_{\nu} = \frac{3 f_L}{3 + a_1} Q p^{3/4} t \propto D^{\left( 8 - 7 \beta \right) / 12}.
\label{adiabat}
\end{equation}
With equations (\ref{length}) and (\ref{pressure}) we can express the source age $t$ as a function of the lobe length and the pressure inside the lobe. The luminosity is proportional to $p^{7/4} D^3$. For our self-similar model the volume of the lobe, $V$, is proportional to $D^3$ and the energy density of the magnetic field inside the lobe is proportional to $p$. Therefore we recover the well-known result that $L_{\nu} \propto u_{\rm B}^{7/4} V$ \citep[e.g.][]{ml94} for the minimum energy condition of a uniform magnetised plasma inside $V$. 

Note also that the luminosity evolution in the adiabatic regime can be positive for $\beta < 8/7$. Sources located in atmospheres with a comparatively flat density distribution {\em increase\/} in luminosity as they grow, if radiative energy losses can be neglected. The spectral index in this regime is $\alpha = 0.5$. 

\subsection{Inverse Compton losses}

The mathematical description of the inverse Compton losses of the relativistic electrons is the same as for synchrotron radiation losses with the energy density of the magnetic field, $u_{\rm B}$, replaced with the energy density of the CMB photon field, $u_{\rm CMB}$ (see equation \ref{dotgamma}). While $u_{\rm B}$ decreases with time, $u_{\rm CMB}$ remains virtually constant over the lifetime of the jet flow. It is therefore inevitable that the inverse Compton losses will eventually dominate over synchrotron losses. In this regime we can neglect $u_{\rm B}$ compared to $u_{\rm CMB}$ and therefore equation (\ref{gamma}) reduces to
\begin{equation}
g \left( x ,  t\right) = \frac{4 \sigma _{\rm T} u_{\rm CMB} t}{3 a_4 m_{\rm e} c}.
\end{equation}
This equation still takes into account adiabatic energy losses. Substituting this into equation (\ref{xmin}) and solving for $x_{\rm min}$ yields
\begin{equation}
x_{\rm min} = \left( 1 - \frac{ 3 a_4 m_{\rm e} c}{4 \sigma _{\rm T} u_{\rm CMB}} \frac{1}{\gamma_{\nu} t} \right)^{1/a_4}.
\label{xminic}
\end{equation}

The time-dependent part of this expression is given by $ 1 / \gamma_{\nu} t$. From equations (\ref{pressure}) and (\ref{gammat}) we find that
\begin{equation}
\frac{1}{\gamma_{\nu} t} \propto t^{ 3 \left( \beta -8\right) / \left[ 4 \left( 5 - \beta \right) \right]}.
\end{equation}
For any reasonable choice of $\beta$, the function $1 / \gamma_{\nu} t$ decreases with increasing $t$. Hence at late times in the source evolution we can use a binomial expansion to simplify equation (\ref{xminic}) and find the luminosity in this regime as
\begin{equation}
L_{\nu} \sim \frac{3 m_{\rm e} c^2 f_n}{14 \sqrt{A} u_{\rm CMB} \nu} Q p \propto D^{\left( -4 -\beta \right) / 3}.
\label{compton}
\end{equation}
The luminosity in this regime always decreases as the source grows. Also note that the spectrum of the source now has a spectral index of $\alpha = 1$. 

\subsection{Synchrotron losses}
\label{syncsec}

As pointed out in the previous section, the energy density of the magnetic field, $u_{\rm B}$, decreases as the source grows. So at early times in the source evolution $u_{\rm B}$ will dominate over $u_{\rm CMB}$ and the luminosity evolution is governed by the synchrotron losses. In the expression for $g \left( x,t \right)$ we now neglect $u_{\rm CMB}$ and so we arrive at
\begin{equation}
x_{\rm min} = \left( 1 - \frac{3 a_3 m_{\rm e} c}{4 \sigma_{\rm T}} \frac{1}{\gamma_{\nu} t u_{\rm B}} \right)^{1/ a_3}
\label{sync}
\end{equation}
for $a_1 \ne 3/5$ and
\begin{equation}
x_{\rm min} = \exp \left( - \frac{3 m_{\rm e} c}{4 \sigma_{\rm T}} \frac{1}{\gamma_{\nu} t u_{\rm B}}\right)
\label{expxmin}
\end{equation}
for $a_1 = 3/5$. For adiabatic and synchrotron losses the time-dependent part of the expression for $x_{\rm min}$ is given by $ 1 / \gamma_{\nu} t u_{\rm B}$. From equations (\ref{pressure}), (\ref{ub}) and (\ref{gammat}) we now find
\begin{equation}
\frac{1}{\gamma_{\nu} t u_{\rm B}} \propto t^{\left( 7 \beta -8\right) / \left[ 4 \left( 5 - \beta \right) \right]}.
\end{equation}
For $\beta > 8/7$ the importance of synchrotron losses decreases as the source grows, while for $\beta < 8/7$ the opposite applies. In environments with a steep density distribution the pressure, and therefore the strength of the magnetic field, inside the lobes decreases fast enough so that the continued injection of freshly accelerated relativistic electrons overcompensates the synchrotron losses. In less stratified atmospheres the synchrotron losses are so severe that the addition of new relativistic electrons cannot compete with them.

Analogous to the discussion in the section above, we can derive an expression for the radio luminosity when the second term in the bracket in equation (\ref{sync}) is small. This limit corresponds to young sources for $\beta > 8/7$ and to older sources for $\beta < 8/7$. In both cases we can use another binomial expansion and get
\begin{equation}
L_{\nu} \sim \frac{m_{\rm e} c^2 f_n}{6 \sqrt{A} \nu} Q = {\rm constant}.
\end{equation}
We arrive at the same result for the case $a_1 = 3/5$ by expanding the exponential function to first order. For $\beta > 8/7$ sources start out at a constant luminosity which is independent of the density parameter $\rho a^{\beta}$, i.e. it is independent of the properties of the environment it is located in. As the source grows, synchrotron losses become less important, the luminosity starts to decrease and the source enters the regime where only adiabatic losses are important. For $\beta < 8/7$ sources start in the adiabatic regime and increase in luminosity until synchrotron losses dominate and the luminosity takes on a constant value, which is again independent of the source environment. 

Whenever the luminosity is dominated by synchrotron losses, the spectral index is again given by $\alpha =1$. 

\subsection{Overall source evolution}

In general the gas density in the environments of radio-loud AGN is not well fitted with a single power-law. A $\beta$-model of the form
\begin{equation}
\rho_x = \frac{\rho}{\left[ 1 + \left( r / a \right)^2 \right]^{\beta /2}}
\label{beta}
\end{equation}
usually fits the X-ray emission from the hot gas in elliptical galaxies, galaxy groups and galaxy clusters \citep[e.g.][]{fmo05}. The lobes of a radio-loud AGN with FRII-type morphology will therefore first expand in an atmosphere with roughly constant density until it reaches a size comparable to $a$. At this point the source evolution changes, since for larger sizes the density of the external medium is approximated by a power-law as given in equation (\ref{exdens}). The evolution of a source in a changing density profile is described in detail in \citet{pa00}. Here we are interested mainly in the overall luminosity evolution of sources as they grow as we will use this information in constructing the radio luminosity function and the distribution of lobe lengths.

It is useful to calculate order-of-magnitude numerical estimates for some of the quantities predicted by the model. For this purpose we set model parameters to physically meaningful, fiducial values. The model input parameters are summarised in table \ref{fidpara}, while model quantities derived from these inputs are collected in table \ref{fidder}.

\begin{table}
\begin{tabular}{lc}
Model parameter & Value\\
\hline
$\Gamma_l$ & 4/3\\
$\Gamma_x$ & 5/3\\
$\beta$ & 0, 2\\
$A$ & 4\\
$\gamma_{\rm min}$ & 1\\
$\gamma_{\rm max}$ & $10^6$\\
$m$ & 2\\
$Q$ & $10^{38}$\,W\\
$\rho$ & $10^{-22}$\,kg\,m$^{-3}$\\
$a$ & 2\,kpc\\
$T$ & $10^7$\,K\\
$u_{\rm CMB}$ & $4\times 10^{-14}$\,J\,m$^{-3}$\\
\hline
\end{tabular}
\caption{Parameters of the fiducial model}
\label{fidpara}
\end{table}

\begin{table}
\begin{tabular}{llc}
Density regime & Parameter & Value\\
\hline
independent & $r$ & 3/4\\
& $p_0$ & $1.4 \times 10^{-11}$\,J\,m$^{-3}$\\
& $f_n$ & $1.2 \times 10^{12}$\,s$^2$\,kg$^{-1}$\,m$^{-2}$\\[2ex]
$\beta =2$ & $c_1$ & 1.5 \\
& $a_1$ & 3/2\\
& $a_3$ & -3/2\\
& $a_4$ & 1/2\\
& $f_p$ & 0.11\\
& $f_{\gamma}$ & $7.2 \left( \frac{\nu}{\rm GHz} \right)^{1/2}$\,J$^{1/4}$\,m$^{-3/4}$\\
& $f_L$ & $3.4 \times 10^{-17}\left( \frac{\nu}{\rm GHz} \right)^{-1/2} $\,J$^{1/4}$\,m$^{1/4}$\,s$^2$\,kg$^{-1}$\\[2ex]
$\beta =0$ & $c_1$ & 1.7\\
& $a_1$ & 3/5\\
& $a_3$ & 0\\
& $a_4$ & 4/5\\
& $f_p$ & 0.11\\
& $f_{\gamma}$ & $7.3 \left(\frac{\nu}{\rm GHz} \right)^{1/2}$\,J$^{1/4}$\,m$^{-3/4}$\\
& $f_L$ & $3.4 \times 10^{-17} \left( \frac{\nu}{\rm GHz} \right)^{-1/2} $\,J$^{1/4}$\,m$^{1/4}$\,s$^2$\,kg$^{-1}$\\
\hline
\end{tabular}
\caption{Model parameters derived for the fiducial model in the density regimes discussed in the text. Note that the value of the derived parameters often does not depend strongly on the value of $\beta$.}
\label{fidder}
\end{table}

In our model all sources start out with an FRII-type morphology. They first propagate through the central regions of their environments where the density is essentially constant. During this {\bf phase 1}, $\beta =0$ and the lobe length increases as  (see equation \ref{length})
\begin{equation}
D = c_1 \left( \frac{Q}{\rho} \right)^{1/5} t ^{3/5}.
\end{equation}

The luminosity evolution of the source in phase 1 is initially governed by adiabatic losses only which implies an increasing luminosity proportional to $D^{2/3}$ (see equation \ref{adiabat}). Synchrotron losses become more important though as the source grows and the luminosity levels out. For the values of our fiducial model we find from equation (\ref{expxmin}) that $x_{\rm min} \sim 0.5$ for $D=a$. Synchrotron losses therefore halve the luminosity of the source compared to the case of pure adiabatic energy losses.

For very small sources we would need to take into account synchrotron self-absorption as demonstrated by the spectral shape of Giga-Hertz Peaked Spectrum (GPS) sources. However, we are mostly concerned with larger sources, at least comparable to the sizes of Compact Steep Spectrum (CSS) objects, and so we neglect self-absorption. 

Note that the radio spectrum in phase 1 is predicted to steepen as the source grows and the luminosity tends to a constant value. Our model therefore predicts that young sources should have relatively steep spectra as confirmed for CSS objects. The steep spectra arise because of the growing importance of synchrotron losses in the comparatively dense core region of the source environments. However, the CSS objects are not `frustrated' by an exceptionally high density. The uniform density of the environment simply limits the expansion speed in such a way as to allow synchrotron losses to accumulate. Even sources in less dense environments will go through this phase. It is the shape of the density distribution rather than the absolute value of the density itself that leads to a steep radio spectrum for CSS objects. 

During phase 1 sources with weaker jets may become turbulent and develop an FRI-type morphology. This is discussed below. Sources retaining laminar jets and their large-scale morphology of type FRII grow eventually larger than the core radius $a$ and enter {\bf phase 2}. The external density distribution is now approximated by a power law. The lobes grow faster than before, but the exact relation between the lobe length $D$ and the source age $t$ depends on the value of the power law exponent $\beta$ (see equation \ref{length}). In phase 2 synchrotron losses become less important as the lobes grow provided that $\beta > 8/7$. The spectrum flattens again and the luminosity evolution follows equation (\ref{adiabat}). 

The last phase, {\bf phase 3}, is reached when the energy density of the magnetic field in the lobe falls below the energy density of the CMB. The lobe continues to expand in the same way as in phase 2, but the luminosity evolution steepens according to equation (\ref{compton}). At the same time the spectrum also steepens again. 

\begin{figure}
\includegraphics[width=8.45cm]{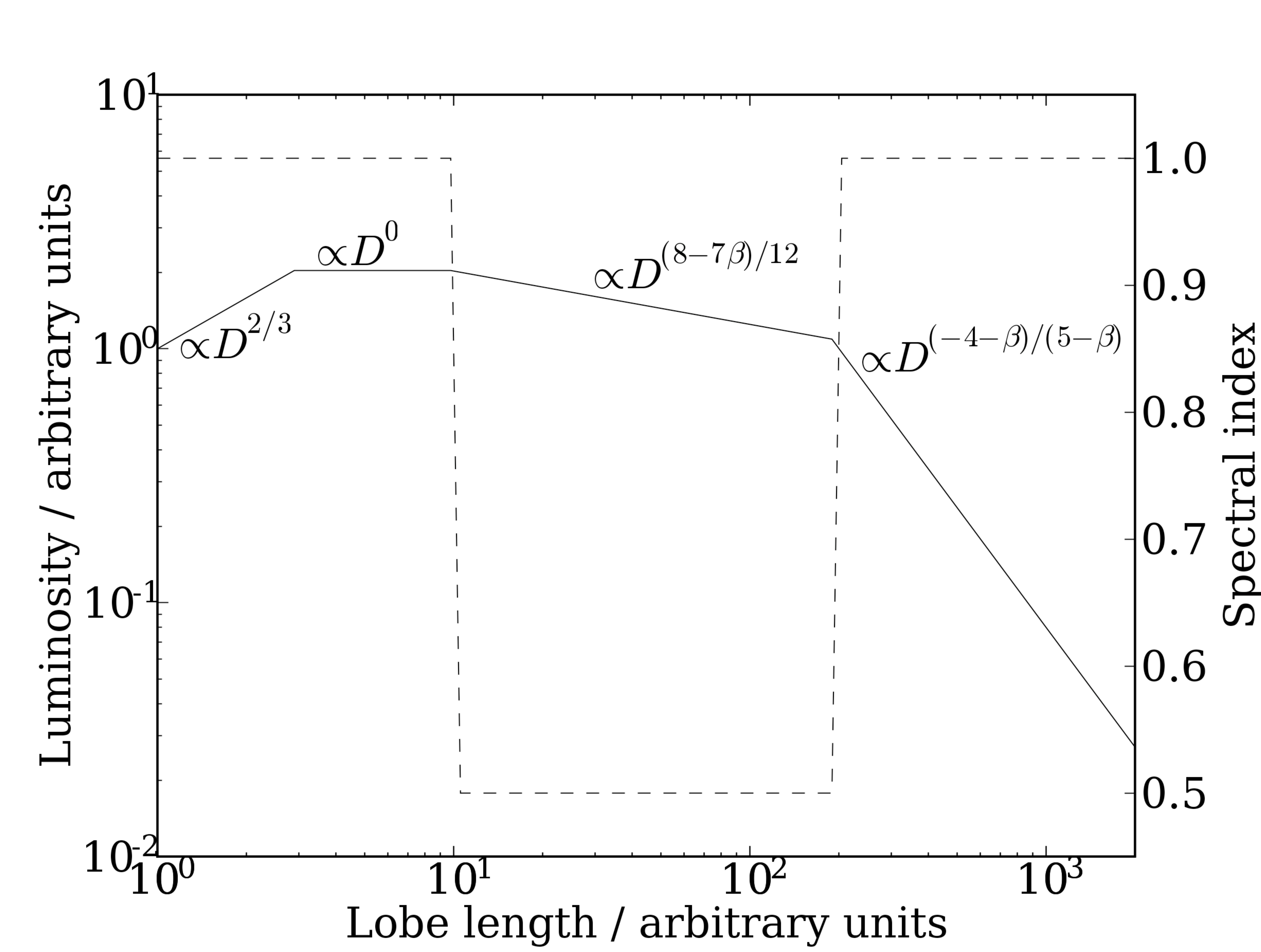}
\caption{Sketch of the luminosity (solid line) and spectral index (dashed line) evolution of a lobe with an FRII-type morphology. The transitions from one regime to the other are of course expected to be less sharp than shown here.}
\label{schematic}
\end{figure}

The end of the source lifetime is reached when the jet flow stops altogether. Depending on the fuelling mechanism for the AGN activity and the jet production mechanism this may occur during any of the evolutionary phases discussed above. However, for most powerful jets sustaining an FRII-type morphology a typical jet lifetime of $10^8$\,years allows the source to evolve through all three phases. A schematic representation of the luminosity and spectral index evolution of the lobe as a function of its length is presented in Figure \ref{schematic}.

\section{FRI sources}
\label{modfri}

For the purposes of this paper, we only consider FRI-type sources with turbulent jets. We do not consider FRI-type sources with a `fat double' morphology \citep[e.g.][]{ol89,ow91}. 

\subsection{Transition to an FRI-type morphology for sources with weak jets}
\label{trans}

The development of an FRII-type structure, i.e. a laminar jet flow inside a lobe, depends crucially on the pressure inside the lobe exceeding the pressure in the source environment. The jet propagates through the low density lobe and its turbulent disruption by fluid instabilities is unlikely. When the pressure inside the lobe becomes comparable to the ambient pressure, the lobe surface becomes unstable and the denser gas initially replaced by the lobe expansion starts to refill the lobe volume \citep[e.g.][]{pa02}. For a roughly isothermal external atmosphere with temperature $T$ the pressure distribution has the same profile as the density distribution, i.e.
\begin{equation}
p_x = \frac{p_0}{\left[ 1 + \left( r/ a \right)^2 \right]^{\beta / 2}},
\end{equation}
with $p_0 = k_{\rm B} T \rho / \bar{m}$, where $\bar{m}$ is the mean mass of the gas particles. The pressure is higher closer to the centre of the distribution and so the replacement of the lobe material starts here after pressure equilibrium is reached. This gives rise to the central `pinching' of the lobe observed in some FRII-type sources \citep{jb82}. 

For sources with lobes of length $D<a$, the lobe pressure decreases rapidly, $p\propto D^{-4/3}$, while the pressure of the ambient medium is roughly constant. Once the lobe extends beyond $a$, the pressure of the source environment decreases as $D^{-\beta}$ while the lobe pressure is proportional to $D^{\left( -4 -\beta \right) /3}$. Since for the great majority of sources $1 < \beta \le 2$, the ratio of ambient and lobe pressure remains almost constant. In other words, once a lobe escapes the core region of its environment without reaching pressure equilibrium, it will remain overpressured for the rest of its evolution. 

The dense external gas replacing the lobe will eventually reach the jet flow itself. At this point the protective function of the lobe disappears and the jet can develop instabilities on its surface leading to its turbulent disruption. At this point the source morphology changes from FRII to FRI. The limiting case of a lobe just escaping from the core region and therefore avoiding the disruption of its embedded jet is given by a lobe with a radius $D / A = a$ at the point in time where $p=p_0$. Using equation (\ref{pressure}) we find the limiting jet power
\begin{equation}
Q_{\rm min} = \left( \frac{k_{\rm B}}{\bar{m} f_p} \right)^{3/2} A^{\left(4 + \beta \right)/2} T^{3/2} \rho a^2.
\label{qmin}
\end{equation}
For our fiducial model we derive $Q_{\rm min} = 3.3 \times 10^{37}$\,W, close to the dividing line between the two FR classes found by \citet{rs91} from the observed correlation of radio luminosity and the strength of optical line emission from the AGN itself. Jets with energy transport rates below $Q_{\rm min}$ will start out with an FRII-type morphology, but turn into FRI-type objects once their jets are disrupted. More powerful jets in the same environment retain their FRII-type morphology throughout their lifetime. 

Both the jet power and the properties of the ambient density distribution determine whether a source makes the transition to an FRI-type morphology or not \citep[see also][on this point]{gb94,gw01,pa00}. We therefore expect that sources with jet powers close to $Q_{\rm min}$ and located in environments with an asymmetric density distribution may develop a hybrid morphology. The fact that only few such objects are known, all of which have radio luminosities close to the demarcation line between the FR classes \citep{gw00}, argues for mainly symmetrical source environments.

\subsection{Dynamics and luminosity of an FRI source}
\label{fri}

Turbulent FRI-type jets often appear to emerge as laminar flows from their AGN. They then develop a flare point where the jet suddenly widens. The flare point is usually the brightest feature of the large-scale radio structure in these objects. After passing through the flare point, the jet flow appears turbulent and resembles smoke rising out of a chimney \citep[e.g.][]{mg91}. It seems natural that the jet decelerates strongly in the flare point and that the synchrotron emitting particles are mainly accelerated here \citep{bog97,lb02a,lb02b,lb04}. Jet deceleration and particle acceleration in the flare point strongly suggest an interpretation of this point as a strong shock within the jet flow. 

Models that relate the radio luminosity of the jet to the turbulent jet flow after passing through the flare point are difficult to construct and mainly rely on conservation equations rather than detailed dynamics \citep{gb94,gb95}. It is also possible to infer details of the jet flow from high resolution radio observations of turbulent jets \citep{lb02a,lb02b,lb04}, but a fully analytic model for the luminosity of FRI-type sources based on the jet dynamics equivalent to the model for FRII sources described above and in the appendix is not available. Despite this lack of a detailed model we show below that the lobe size of an FRI-type source is not well defined, while its observed radio luminosity is likely to be constant for most of its lifetime. 

In our model for sources with laminar jets inflating FRII-type lobes, we assume that the jet is ballistic when it emerges from the AGN and comes into pressure balance with its lobe by passing through a reconfinement shock \citep[see][KA]{cr91,sf91}. After the external medium has replaced the lobe material in sources with weak jets, the initially ballistic jet still needs to adjust to the pressure now exerted on it by the external medium. As the bulk velocity of the jet is supersonic with respect to the jet internal sound speed, this adjustment still causes a shock in the jet flow. In this reconfinement shock a fraction of the kinetic energy of the jet is dissipated to thermal energy. If the fraction is large, then the jet is even more susceptible to turbulent disruption downstream of the shock. For a fixed pressure in the external gas, weaker jets are more prone to disruption than more powerful jets as they have to dissipate a larger fraction of their power to reach pressure equilibrium. We identify the flare point observed in FRI-type jets with the reconfinement shock \citep{lb02b}.

More powerful jets may avoid turbulent disruption, even when in contact with the dense, external medium. However, usually the lobes inflated by these more powerful jets do not come into pressure equilibrium with the external gas in the core while the jet flows are still young. The jets can then inflate a large lobe, but if $\beta < 2$, then the lobe pressure will eventually fall below the pressure in the ambient medium and pinching of the lobes at their centres can occur \citep{hmw94}. In this case the eventual replacement of the lobe may lead to the formation of a second, small lobe of type FRII closer to the host galaxy than the previously inflated lobe. This may explain the Double-Double Radio Galaxies (DDRGs) first described by \citet{sbrlk00}, which always consist of very large (Mpc-scale) outer lobes and much smaller inner lobes. 

After passing through the reconfinement shock the bulk velocity of weak jets will be slow. The jets inflate turbulent lobes in which the jet material and the external medium will start mixing. This process gives rise to the similarity of the lobes and smoke from a chimney. For very slow jets the analogy may extend further as buoyancy forces and bulk motion in the external medium may play an important role in shaping the lobes. Good examples for the latter effect are the narrow-angle tail sources where the host galaxies moves through the external medium creating wake-shaped radio lobes \citep[e.g.][]{oo86}. The turbulent distortions of the lobe combined with the progressive mixing of the jet material with the external gas, make it difficult to define the lobe size using radio observations. Furthermore, the lobes of FRI-type objects fade from the flare points outwards and only very deep radio observations can map a significant fraction of their total extent \citep[e.g.][]{br01}. 

The flare point is usually the brightest feature of an FRI-type source. The relativistic electrons emitting synchrotron radiation are mostly accelerated here. If we assume that a fixed fraction of the pressure inside the jet is provided by the magnetic field and the synchrotron emitting relativistic electrons, then the radio luminosity of the lobe with volume $V$ is roughly given by
\begin{equation}
L_{\nu} \propto p_0^{7/4} V.
\end{equation}
As discussed above, the size of the lobe and therefore its volume cannot be easily established from observations. Concentrating on the adiabatic inflation of the lobe immediately downstream of the flare point where the external pressure is roughly constant, we get from energy conservation
\begin{equation}
{\rm d} U = \left( \Gamma_j -1 \right) p_0 \, {\rm d} V = - p_0 \, {\rm d} V + Q \, {\rm d} t,
\end{equation}
where $\Gamma _j$ is the adiabatic index of the jet material. After a time $t$, the lobe has a volume
\begin{equation}
V = \frac{Qt}{\Gamma _j p_0},
\end{equation}
and the radio luminosity of the lobe is $L_{\nu} \propto p_0^{3/4} Q t$. This result neglects any energy losses of the electrons and does not take into account the drop in pressure in the environment once the lobe extends beyond the core radius $a$. Both effects are responsible for the diminishing brightness of the lobe at increasing distances from the flare point. In practice, for a given surface brightness sensitivity these outer structures will contribute a decreasing fraction of the overall luminosity for increasing distance from the flare point.

In our simplified model the dimming of the outer regions of the lobe puts an upper limit $t_{\rm lim}$ on $t$, the time during which the observable lobe was inflated. If we observe the lobe at time $t$, then those parts inflated at times $< t- t_{\rm lim}$ will not contribute to the luminosity. The limit $t_{\rm lim}$ depends on the pressure and the strength of the magnetic field in the lobe. Hence it depends on the pressure distribution of the external medium, but not on the jet power. Once the lobe has reached a volume $V = Q t_{\rm lim} / \left( \Gamma_j p_0 \right)$, its observable radio luminosity should remain roughly constant in this simple picture. Note that its total luminosity may in principle continue to slowly increase, but this is not detectable in an observation with a given surface brightness limit. Note also that the observable luminosity is linearly proportional to the jet power $Q$.

Clearly the ideas outlined above do not consitute a proper model of sources with an FRI-type morphology. For example, we neglect the possibility of particle acceleration in the turbulent lobes. Nevertheless, based on the discussion above, in the following we will use the reasonable results that the observable radio luminosity is roughly constant and linearly proportional to the jet power. 

\section{The radio luminosity function (RLF)}
\label{rlf}

The RLF of extragalactic radio sources is commonly parameterized as a broken power-law of the form 
\begin{equation}
n(L_{\nu}) = n_0 \left[ \left( \frac{L_{\nu}}{L_*} \right)^{\zeta_{\rm low}} + \left( \frac{L_{\nu}}{L_*} \right)^{\zeta_{\rm high}} \right]^{-1},
\end{equation}
where $\zeta_{\rm  low}$ and $\zeta_{\rm high}$ are the power-law exponents at low and high luminosities respectively. The transition between these two regimes occurs at the break luminosity $L_*$. For convenience, the RLF is usually expressed as number density of sources per unit logarithm of the radio luminosity. We follow this convention. 

The break luminosity is close to the luminosity separating the two FR classes with FRI-type sources dominating below the break and FRII types mainly found at luminosities above $L_*$ \citep{fr74}. Note however that the luminosity separating the two classes appears to depend on the optical luminosity of the host galaxy \citep{lo96}. For now we assume that the break in the RLF at $L_*$ is fixed and that the slopes of the RLF below and above the break, $\zeta_{\rm low}$ and $\zeta_{\rm high}$, are determined only by FRI-type and FRII-type sources respectively. We will return to the more complicated, observed separation of the two classes in terms of radio and optical luminosity further on.

\subsection{The low luminosity end of the RLF}
\label{low}

We first concentrate on the low luminosity end of the RLF where $n \propto L_{\nu}^{-\zeta_{\rm low}}$. Above we suggest that sources with an FRI-type morphology evolve out of small FRII structures for sources with weak jets. Let $S(L_{\nu})$ be the rate per unit volume and per unit luminosity at which sources with FRI-type morphology and luminosity $L_{\nu}$ appear in the universe. The time it takes sources described by our fiducial model parameters, but with a jet power equal to $Q_{\rm min}$ to evolve through their initial FRII stage is $5\times 10^6$\,years. Most sources that disrupt will have jets less powerful than this and therefore disrupt earlier. This is short compared to their total lifetime of about $10^8$\,years. We therefore expect $n (L_{\nu}) L_{\nu}^{-1} \propto S(L_{\nu})$. The factor $L_{\nu}^{-1}$ arises from the definition of $n(L_{\nu})$ as the density of sources per unit logarithm of $L_{\nu}$. 

We argued in section \ref{fri} that the radio luminosity of sources after they develop an FRI-type morphology is constant and linearly proportional to the jet power $Q$. We can therefore relate $S(L_{\nu})$ to the rate per unit volume and per unit jet power at which AGN producing jets with a given $Q$ are appearing in the universe, $S'(Q)$, as $S'(Q) \propto S(L_{\nu})$. Therefore the `brith rate' of jets from AGN with a jet power $Q$ must be $S'(Q) \propto Q^{-1-\zeta_{\rm low}}$. 

There is no {\em a priori}\/ reason to assume that this form of $S'(Q)$ also applies at the higher jet powers of the sources dominating the RLF above $L_*$. However, in the following we will show that this assumption leads to a self-consistent explanation of the shape of the RLF. For further discussion of this point see sections \ref{mode} and \ref{micro}.

\subsection{The high luminosity end of the RLF}
\label{high}

At luminosities $L_{\nu} > L_*$ the RLF is dominated by powerful sources with FRII-type morphologies. For these we have to take into account the luminosity evolution of their lobes. Consider for now sources with a luminosity evolution described by $L_{\nu} \propto t^{-\delta}$, where $\delta$ will take different values according to which regime the source is currently in. The evolution of the RLF above $L_*$ is then given by \citep[e.g.][]{cmw71}
\begin{equation}
\frac{\partial N}{\partial t} +\frac{\partial}{\partial L_{\nu}} \left( N \frac{\partial L_{\nu}}{\partial t} \right) = S (L_{\nu}),
\label{difrlf}
\end{equation}
where $N$ is the number density of radio sources per unit luminosity and so $n = \ln (10) L_{\nu} N$. 

We assume that the RLF changes only on cosmological timescales, i.e. on timescales much longer than the lifetime of individual sources. This allows us to set $\partial N / \partial t = 0$. The solution of the differential equation is then given by
\begin{equation}
N \propto L_{\nu}^{- \left( \delta + 1 \right) / \delta} \int_{L_{\nu}}^{L_{\rm max}} S \left( L_{\rm peak} \right) \, {\rm d} L_{\rm peak}.
\label{densinte}
\end{equation}
We discuss the various terms in the following.

Sources retaining an FRII-type morphology initially expand in the core region of the ambient gas distribution, $D < a$, where the gas density is roughly constant. The time taken to traverse this core region will be short compared to the total lifetime of the source. Therefore we simplify the construction of the RLF by assuming that sources only start to contribute to the high luminosity end of the RLF once their lobes extend beyond the core region, i.e. when $D=a$. The discussion in section \ref{syncsec} implies that the radio luminosity in this phase of the source evolution tends to a constant $L_{\rm peak}$ and $L_{\rm peak} \propto Q$. In some sources synchrotron losses may not become significant during this phase. For these sources equation (\ref{adiabat}) implies that $L_{\rm peak} \propto Q^{7/6}$ for $D=a$, very close to the result for a constant luminosity. In the following we use $L_{\rm peak} \propto Q$ for all sources. 

Once the lobes extend beyond the core region, their luminosity evolution is described by equation (\ref{adiabat}) and so $\delta = \left( 7 \beta -8 \right) / \left[ 4 \left( 5 - \beta \right) \right]$. As long as $\beta > 8/7$, the luminosity
of the source always decreases. Hence the luminosity $L_{\rm peak}$ at $D=a$ is indeed the maximum luminosity of the source at any time during its life. The integration of the source function in equation (\ref{densinte}) sums up the contribution to the RLF of all sources with $L_{\rm peak} \ge L_{\nu}$.

Most sources retaining an FRII-type morphology will reach within their lifetime the regime where inverse Compton losses of the electrons in the lobe become important. In this regime the luminosity evolution accelerates as $\delta$ changes to $\delta' = \left( 4 + \beta \right) / \left( 5 - \beta \right)$ (see equation \ref{compton}). We will show below that sources in the inverse Compton regime do not make a significant contribution to the RLF. Hence the upper limit of the integration is set by $L_{\rm max} = L_{\rm peak}$, where $L_{\rm peak}$ is taken for those sources which enter the inverse Compton regime when their luminosity is equal to $L_{\nu}$.

Using the results from above that $L_{\rm peak} \propto Q$ and $S \left( L_{\rm peak} \right) \propto S' \left( Q \right) \propto Q^{-1-\zeta_{\rm low}}$ we can solve the integral,
\begin{equation}
N \propto L_{\nu}^{-\left( \delta + 1 \right) / \delta} \left( L_{\nu}^{-\zeta_{\rm low}} - L_{\rm max}^{-\zeta_{\rm low}} \right) \sim L_{\nu}^{-\left( \delta + 1 \right) / \delta -\zeta_{\rm low}},
\end{equation}
where for the last step we assume $L_{\rm max} \gg L_{\nu}$. $L_{\rm peak}$ is the maximum luminosity of a source at any point during its lifetime at a time when $D \sim a$ and $L_{\rm max}$ is equal to $L_{\rm peak}$ for the sources with the most powerful jets contributing to the RLF at $L_{\nu}$. Their luminosity decreases from $L_{\rm max}$ proportional to $D^{-\delta \left( 5 - \beta \right) / 3}$. These powerful sources will enter the regime dominated by inverse Compton losses only once their lobes have grown in size well beyond $a$. At this point their luminosity has decreased significantly compared to $L_{\rm max}$. 

In principle there is a second contribution to $N$ from sources already in the regime dominated by inverse Compton losses. It is easy to see that this additional contribution has the form
\begin{equation}
N_{\rm IC} \propto L_{\nu}^{- \left( \delta' + 1 \right) / \delta'} \left( L_{\rm max}^{-\zeta_{\rm low}} - L_{\rm lim}^{-\zeta_{\rm low}} \right),
\end{equation}
where $L_{\rm lim} > L_{\rm max}$ is now the maximum luminosity of any source at any time. Comparing with the term describing the sources in the adiabatic regime, we can neglect this contribution to the RLF at $L_{\nu}$ as long as $L_{\rm max} \gg L_{\nu}$.

With the definitions of $n$ and $N$ we now have 
\begin{equation}
n \propto L_{\nu}^{-\zeta_{\rm high}} \propto L_{\nu} N \propto L_{\nu}^{1-\left( \delta + 1 \right) / \delta - \zeta_{\rm low}},
\end{equation}
or, equivalently
\begin{equation}
\zeta_{\rm high} - \zeta_{\rm low} = \frac{1}{\delta} = \frac{4 \left( 5 - \beta \right)}{7 \beta - 8}.
\end{equation}
\citet{wrb01} construct the RLF from observed complete samples of extragalactic radio sources. They find $\zeta_{\rm low} \sim 0.6$ and $\zeta_{\rm high} \sim 2.4$ which is consistent with $\beta \sim 2$. The mean value of $\beta$ for the exponent of the external density distributions of the gaseous haloes of elliptical galaxies, galaxy groups and clusters in the extensive sample of \citet{fmo05} is 1.6 which would imply a larger difference $\zeta_{\rm high} - \zeta_{\rm low}$. However, they also find that $\beta$ is correlated with the temperature of the X-ray emitting gas in the sense that hotter systems have steeper density profiles. The RLF we study here is composed of the most luminous radio sources in the universe, which therefore are preferentially located in hotter and denser systems with steeper density profiles. 

\subsection{The break in the RLF, $L_*$}
\label{break}

Our model can also provide an estimate for the location of the break in the RLF. Consider the FRII-type sources with the weakest jets. Their lobes just escape the core region without suffering turbulent disruption. Their radio luminosity then follows the relation given in equation (\ref{adiabat}) until the energy density inside their lobes becomes comparable to the energy density of the CMB, $u_{\rm CMB}$. The source then enters the inverse Compton dominated regime and its luminosity evolution steepens. From equation (\ref{magdens}) it follows that at this time $p= 7 u_{\rm CMB} / 12$ and using equation (\ref{pressure}) the source age is 
\begin{eqnarray}
t_{\rm IC} & = & \left( \frac{12 f_p}{7 u_{\rm CMB}} \right)^{\left( 5 - \beta \right) / \left( 4 + \beta \right)} c_1^{\left( \beta - 5 \right) / 3} \nonumber\\
& \times & \left( \rho a^{\beta} \right)^{ 3 / \left( 4 + \beta \right)} Q_{\rm min} ^{\left( 2 -\beta \right) / \left( 4 + \beta \right)}.
\end{eqnarray}
The luminosity of the lobe at $t_{\rm IC}$ is found from equation (\ref{adiabat}) as
\begin{eqnarray}
L_{\rm IC} & = & \frac{3 f_L}{3 + a_1} c_1^{\left( \beta - 5 \right) / 3} \left( \frac{7 u_{\rm CMB}}{12} \right)^{\left( 7\beta - 8 \right)/ \left[4 \left( 4 + \beta \right) \right]} \nonumber\\
& \times & \left( \rho a^{\beta} \right)^{3 / \left( 4 + \beta \right)} Q_{\rm min}^{6 / \left( 4 + \beta \right)}.
\label{lic}
\end{eqnarray}
Note that for $\beta \sim 2$, $L_{\rm IC}$ is roughly linearly proportional to $Q_{\rm min}$ while $t_{\rm IC}$ is independent of the jet power. All sources in comparable environments enter the inverse Compton dominated regime at about the same age, independent of their jet power. However, sources with $Q=Q_{\rm min}$ have the lowest radio luminosity and remain the least luminous objects at $t>t_{\rm IC}$.

After entering the inverse Compton dominated regime the radio luminosity of the lobe decreases further according to equation (\ref{compton}) with $L_{\nu} \propto t^{\left(-4-\beta \right) / \left( 5 - \beta \right)}$. Therefore, if the jet shuts down at a time $t_{\rm final} \sim 10^8$\,years, we have
\begin{equation}
L_{\rm final} = L_{\rm IC} \left( \frac{t_{\rm final}}{t_{\rm IC}} \right)^{\left(-4 - \beta \right)/ \left( 5 - \beta \right)}.
\end{equation}
We expect the break in the RLF to occur between $L_{\rm final}$ and $L_{\rm IC}$, i.e. $L_{\rm final} \le L_* \le L_{\rm IC}$. Also, variations in the properties of the external density profiles will tend to further broaden the range in luminosity over which the RLF changes its slope. For our fiducial model at $\nu=151$\,MHz, with $Q_{\rm min} = 3.3\times 10^{37}$\,W derived above and $t_{\rm final} =10^8$\,years, we find $t_{\rm IC} = 2.8 \times 10^7$\,years, $L_{\rm final} = 2.3 \times 10^{25}$\,W\,Hz$^{-1}$ and $L_{\rm IC} = 2.9 \times 10^{26}$\,W\,Hz$^{-1}$ in agreement with the break of the RLF derived from observations \citep{wrb01}.

\section{The size distribution of FRII-type sources}
\label{sizeII}

The size of lobes with an FRII-type morphology increases with lifetime of the jet flow. However, factors like the jet power and the density distribution of the external medium influence the rate of growth (see equation \ref{length}). Studying the size distribution of FRII-type lobes in flux-limited samples in detail is further complicated by the possible cosmological evolution of these quantities as well as the overall lifetime of the jets and the selection effects introduced by the flux limit in the cosmological context \citep[e.g.][]{ner95,ka98a,br99,wdg00,bw06,bw07}. Here we restrict the discussion to a very general study of the size distribution predicted by our model without taking into account cosmological effects. 

\citet{ob97} point out that compact sources with sizes below a few kpc are so numerous in observed samples, that not all of them can evolve into sources with lobes tens and hundreds of kpc across. The number of sources per unit logarithm of lobe length, $n(D)$, is constant below a few kpc, while $n (D) \propto D^{0.4}$ for larger $D$. These findings can be explained with an intermittently active jet flow \citep{rb97}, but here we explain the same result with our model including a continuous jet flow. Our approach is similar to that of \citet{pa00} who also invokes jet disruption of small sources to explain the large number of small compared to large sources. 

\subsection{Large sources}

We can use our model to predict $N(D)$, the number of sources per unit lobe length, using a formalism analogous to the RLF. As before, $n(D)$ is related to $N(D)$ as $n(D) = D N(D)$. Neglecting the possible cosmological evolution of $N(D)$, we can write (compare to equation \ref{difrlf} for the RLF)
\begin{equation}
\frac{\partial }{\partial D} \left( N \frac{\partial D}{\partial t} \right) = - H(D),
\label{size}
\end{equation}
where $H(D)$ is a sink function. As sources grow, their radio luminosity usually decreases and so at some point they drop below the flux limit of the sample used to construct $N(D)$ from observations. This effect is represented by $H(D)$.

The length of the lobe as a function of time is given by equation (\ref{length}) as $D \propto t^{3 / \left( 5 - \beta \right)}$. The solution for the size distribution is then
\begin{equation}
N(D) \propto D^{ \left( 2 - \beta \right) / 3} \int_D^{D_{\rm min}} H(D') \,{\rm d} D'.
\end{equation}
The integration of the sink function simply sums up all sources that have dropped below the flux limit of the sample when their lobes had lengths shorter than $D$.  Neglecting cosmological effects, the flux limit of the sample translates into a limit on the source luminosity. Sources with comparatively weak jets will drop below this limiting luminosity while they are in the adiabatic regime. Equation (\ref{adiabat}) then implies that the size of the lobe at the point the source drops out of the sample is $D \propto Q^{14 / \left( 7 \beta - 8 \right)}$. To solve the integral, we note that $H(D) \,{\rm d} D \propto S'(Q) \,{\rm d} Q$ and so
\begin{equation}
N(D) \propto D^{ \left( 2 - \beta \right) / 3-\zeta_{\rm low} \left( 7\beta - 8 \right)/14}.
\end{equation} 
With $\beta \sim 2$ and $\zeta_{\rm low} \sim 0.6$, our model then predicts $n(D) \propto D^{0.7}$, somewhat steeper than the observations suggest. For $\beta = 1.6$, the mean of the \citet{fmo05} sample, we find $n(D) \propto D^1$, considerably steeper than the observational value. However, sources with more powerful jets will enter the inverse Compton dominated regime before dropping out of the sample. In this case from equation (\ref{compton}) we have $D \propto Q^{\left( 4 + \beta \right) / 5}$ and so
\begin{equation}
N(D) \propto D^{ \left( 2 - \beta \right) / 3 - \zeta_{\rm low} \left( 4 + \beta \right) /5},
\end{equation}
or $n(D) \propto D^{0.3}$ for $\beta =2$ and $n(D) \propto D^{0.5}$ for $\beta =1.6$. For a sample composed of sources from both regimes, the model therefore results in a slope close to that derived from observations. 

For very large lobe sizes the fact that jets will stop to supply energy to the lobes at $t=t_{\rm final}$ will become relevant. If $t_{\rm final}$ is similar for all sources, then equation (\ref{length}) relates the final length of the lobe to the jet power as $D \propto Q^{1 / \left( 5 - \beta \right)}$. We then find $n(D) \propto D^{-0.8}$. The data point containing the largest sources in \citet{ob97} may indicate this predicted drop in the size distribution.

\subsection{Small sources}

At small lobe sizes we expect to find many objects that currently have an FRII-type morphology, which will develop turbulent jet flows at a later stage. From the discussion of the transition from FRII to FRI morphology in section \ref{trans}, it is clear that for most sources the transition will occur while they are contained within the core radius of the external density distribution, $a$. The density distribution quickly steepens outside $a$ and as discussed above, it becomes easier for sources to avoid disruption in more stratified environments. From these considerations it follows that we expect the change in the slope of $n(D)$ to occur around $D=a$. Given that $n(D)$ as determined from observations has its break at a few kpc, this is consistent with the results of X-ray observations of the gas in elliptical galaxies \citep[e.g.][]{fmo05}. 

The density distribution given by equation (\ref{beta}) can be approximated by power-laws with an exponent changing as a function of distance from the centre. Hence we expect that the jets in most sources become turbulent when their environment is described by a power-law with an exponent $\beta _{\rm turb} < \beta$. Furthermore, at $r=a$ the exponent of the approximating power-law is $\beta / 2$. If sources mostly disrupt inside $a$, we can further constrain the relevant power-law exponent to $\beta_{\rm turb} < \beta/2$. Also, during phase 1 of the source evolution the radio luminosity increases until reaching a constant maximum. The sample of \citet{ob97} contains only very luminous objects. Hence many objects with weak jets disrupt before reaching a luminosity above the sample limit. Only sources with powerful jets are included in the sample and this happens only once they have evolved long enough so that their luminosity exceeds the sample limit. If the jets in these objects subsequently disrupt, they must do so at a time when the lobes have grown to sizes comparable to $a$. Hence we expect $\beta_{\rm turb} \sim \beta / 2$. With this constraint in mind we can construct $n(D)$ from our model for small sources.

We can again use equation (\ref{size}), but this time $H(D)$ describes sources whose jets are disrupted and which develop an FRI-type morphology. The condition for disruption is that the pressure inside the lobe becomes comparable to that in the source environment. From equation (\ref{pressure}) we then find that the lobe size at which disruption occurs is given by $D \propto Q^{2 / \left( 4 + \beta_{\rm turb} \right)}$. The solution of the differential equation is then
\begin{equation}
N(D) \propto D^{\left( 2 - \beta_{\rm turb} \right) / 3 -\zeta_{\rm low} \left( 4 + \beta_{\rm turb} \right) /2}.
\end{equation}
For $n(D) \sim {\rm constant}$ in this regime, we require $\beta_{\rm turb} \sim 0.7$, which is indeed smaller than $\beta / 2 \sim 1$. For $\beta =2$, the density distribution is approximated by a power-law with exponent $\beta_{\rm turb}=0.7$ at a distance of $r\sim 0.7 a$ from the centre. 

In the framework of the model the flat shape of $n(D)$ for small lobes is caused by many small objects developing an FRI-type morphology before growing to large sizes. In section \ref{break} we have shown that the break in the RLF is located around $10^{26}$\,W\,Hz$^{-1}$ at 151\,MHz. Therefore not many sources with an FRI-type morphology can contribute to the RLF at luminosities in excess of this threshold. However, the small sources discussed in \citet{ob97} have luminosities in the range $10^{26}$\,W\,Hz$^{-1}$ to $10^{28}$\,W\,Hz$^{-1}$ at 5\,GHz which, for a spectral index of $\alpha = 0.75$, roughly translates to $10^{27}$\,W\,Hz$^{-1}$ to $10^{29}$\,W\,Hz$^{-1}$ at 151\,MHz. While some sources with an FRI-type morphology and a radio luminosity in excess of $5\times 10^{26}$\,W\,Hz$^{-1}$ at 151\,MHz exist \citep[e.g.][]{lo96,hec07}, they are rare. Therefore our model requires that the radio luminosity of most sources developing FRI-type lobes must drop by at least two orders of magnitude once their jets become turbulent. Without a detailed model for the FR transition it is impossible to show whether this is realistic or not. However, we note that this requirement may make our explanation for the small lobe end of $n(D)$ problematic. \citet{pa00} also suggests that the luminosity of sources with disrupted jets decreases dramatically. However, in that model sources with disrupted jets are not expected to evolve into FRI-type objects, but fade so strongly that they become virtually undetectable. 

\section{Connection to other wavelengths and microquasars}
\label{con}

\subsection{Host galaxies}

In the previous sections we concentrated exclusively on the radio properties of radio galaxies and radio-loud quasars. However, it is well known that the two FR classes also differ in respects other than their radio properties. \citet{lo96} show that the dividing line in terms of radio luminosity between the FR classes is a function of the B-band luminosity of the host galaxy. In our model the dividing luminosity between the FR classes is identified with the break in the RLF, $L_*$. From observations we therefore expect that $L_* \propto L_{\rm B}^{1.8}$, but note that \citet{bkh05b} do not find a dependence of $L_*$ on $L_{\rm B}$. 

In our model $L_*$ is determined by the luminosity of sources with the weakest jets entering the inverse Compton regime, $L_{\rm IC}$, and their luminosity when their jets cease to supply energy to the lobes, $L_{\rm final}$. Combining equations (\ref{qmin}) and (\ref{lic}) we find $L_{\rm IC} \propto \rho^{3/2} a^3 T^{3/2}$ for $\beta =2$. $L_{\rm final}$ shares the same parameter dependence and hence $L_* \propto \rho^{3/2} a^3 T^{3/2}$. The same parameters describing the density distribution in the source environment also determine the X-ray luminosity of the ambient gas. Bremsstrahlung emission depends strongly on the density of the radiating gas and so we expect the total X-ray luminosity of the source environment to be dominated by the contribution of the core region. In this case $L_{\rm X} \propto \rho^2 a^3 \sqrt{T}$. At least in elliptical galaxies the X-ray luminosity is correlated with the optical luminosity of the galaxy as $L_{\rm B} \propto L_{\rm X}^{1/2}$ \citep{ofp01} and so we expect $L_{\rm B} \propto \rho a^{3/2} T^{1/4}$. The temperature of the X-ray emitting gas does not vary strongly between individual objects \citep[e.g.][]{fmo05}, but the core radius and the central density can be different. If the central density $\rho$ was fixed for all galaxies hosting powerful jets  and only the core radius is varying, then we would expect $L_* \propto L_{\rm B}^2$. In the case of a fixed core radius $a$ and a variation in the central density, the expectation would be $L_* \propto L_{\rm B}^{3/2}$. The observed correlation lies between these possibilities and our model is thus consistent with observations \citep[see][for a similar argument]{gw01}. 

\subsection{Nuclear emission}

Differences between the FR classes are also found in the optical emission of the jet-producing AGNs themselves. Most FRII-type sources show strong optical line emission while FRI-types only possess weaker lines \citep[e.g.][]{hl79}. FRI-type objects and some FRII-types with weak emission lines are collectively referred to as Low Excitation Radio Galaxies (LERG) while the objects with strong emission lines are called High Excitation Radio Galaxies (HERG). We emphasise that the FR classes cannot be identified directly with the LERG and HERG groups. There are a number of objects with weak emission lines, but FRII-type radio morphologies \citep{pb89}, and a few objects showing strong emission lines combined with an FRI-type morphology \citep[e.g.][]{br01}. 

The optical continuum emission of LERGs is correlated with the luminosity of their radio cores and is consistent with originating in the inner jet flow rather than the AGN itself \citep{ccc00}. In HERGs the optical continuum luminosity exceeds the value predicted by this correlation for a given radio luminosity implying a contribution by the AGN itself.  These results can be interpreted as evidence for a radiatively inefficient accretion flow in LERGs compared to more efficient flows in HERGs \citep{fkm04}. This interpretation is also consistent with the observed differences between the two classes in X-rays \citep{hec06} and infrared emission \citep{mhs04}.

In our model, we can combine the emission properties of the AGN with the radio morphology classification, if we assume that the radiatively inefficient accretion mode is associated with the production of weak jets. The main ingredient determining whether or not a jet disrupts and forms a lobe with an FRI-type morphology is the jet power. Hence with the above assumption we would expect AGN with radiatively inefficient accretion. i.e LERGs, to produce only weak jets which are more easily disrupted and give rise to radio lobes with an FRI-type morphology. At the same time, HERGs should produce powerful jets which retain their FRII-type morphology. However, as we have seen above, the transition in radio morphology of AGNs is also influenced by the properties of the gaseous halo of the host galaxy. Unless the jet power associated with the change from radiatively efficient to inefficient accretion is for some unknown reason fine-tuned to exactly coincide with the jet power connected to the change in radio morphology, we would expect to observe a number of hybrid objects. We have already mentioned that some LERGs possess FRII-type radio lobes and it is interesting to note that their radio luminosities are close to $L_*$ \citep{pb89}. Furthermore, their precarious position between laminar jet flow and turbulent disruption of the jets may give rise to their prominent jets and weak radio hotspots \citep{hap98}. Finally, HERGs with FRI-type radio lobes appear to be rare \citep[e.g.][]{br01}, suggesting that the jet power associated with the switch 
from radiative inefficient to efficient accretion is lower than the $Q_{\rm min}$ for typical radio source 
environments.

\subsubsection{Fuelling the AGN}
\label{mode}

Recently \citet{hec07} suggested that the accretion flows in LERGs are fundamentally different from those in HERGs. In this scenario LERGs are fuelled by comparatively hot, but radiatively inefficient gas which is cooling out of the galactic atmosphere or the surrounding group or cluster environment. The fuelling by hot gas cooling out of the galaxy atmosphere allows for a feedback loop regulating the cooling of gas and therefore limiting the growth of the host galaxy \citep{bkh06}. In contrast, HERGs are powered by the accretion of cold gas, most likely acquired in a merger of the host galaxy with a gas-rich companion. The latter accretion mode releases more energy in the form of radiation and, in our model, produces more powerful jets than the former. 

The idea of different accretion modes giving rise to different radio morphologies is interesting, but it also creates a problem for our construction of the RLF. The mechanisms for fuelling of sources are quite different in the two accretion modes. The switch between the modes occurs near to $L_*$ and so it is not clear whether the birth rate of jets, $S'(Q)$, determined from the low luminosity end of the RLF in section \ref{low}, can be extrapolated to the high luminosity end as we have done in section \ref{high}. The overlap in radio luminosity of LERGs and HERGs around $L_*$ makes a clear separation of $S'(Q)$ for the two classes very difficult. However, the model is self-consistent and produces the break in the RLF without invoking a difference in the birth functions of LERGs and HERGs. 

\subsubsection{Relation to microquasars}
\label{micro}

An alternative to the above scenario is provided by analogy to microquasars. These objects show radiatively efficient and inefficient accretion states without the need for a difference in the fuel source. The transitions are moderated by the accretion rate through the inner parts of the disc. Instabilities within the disc can lead to large changes of this parameter \citep[e.g.][]{fkr92}. For high accretion rates (high-soft state) the discs are luminous, but formation of jets appears to be suppressed, while for small accretion rates the discs are radiatively inefficient (low-hard state) and comparatively weak jets are produced \citep{fbg04}. During the transition from the low-hard to the high-soft state, powerful, but short lived jet ejections occur. The LERGs may be identified with the low-hard state of microquasars as they also show radiatively inefficient accretion discs and weak jets \citep{kjf06}. In this scheme, the HERGs must be identified with the transition between the states as the formation of powerful jets and a luminous accretion disc is required for them. Radio-quiet quasars may then be the counterparts to objects in the high-soft state.

Is the state transition slow enough in AGN to explain the lifetime of jet flows? It is difficult to measure the duration of the state transition in microquasars, but from simultaneous X-ray and radio observations the relevant timescales appear to be of order several hours to a few days \citep[e.g.][]{fgmmpssw99}. This must be the timescale for the disc instability causing the state transition. It is by no means clear which physical mechanism is responsible for the disc instability. However, it is likely that the relevant process is connected with the viscous timescale within the accretion disc which roughly scales as the mass of the central black hole to the power 3/2 \citep{fkr92}. The mass of the black hole in a typical microquasar is 10\,M$_{\odot}$, while that of a black hole at the centre of an AGN is about $10^8$\,M$_{\odot}$. Hence we would expect the duration of the state transition in an AGN to be of order $10^8$\,years, comparable to the maximum lifetime of jet flows in powerful radio galaxies \citep{al87}.

If this scenario is correct, then the difference in the radiative properties between LERGs and HERGs, and ultimately that between the FR classes, is not caused by differences in the fuelling mechanism. In this case, weak and powerful jets can originate in the same system at different times. The extrapolation of the birth rate of jets $S'(Q)$ to higher jet powers is then less problematic. Also, black holes of a given mass can give rise to jets with a range of jet powers, translating to different radio luminosities of their lobes, at different points during their accretion history. This predicted statistical independence of the radio luminosity and the black hole mass is consistent with observations \citep{bkh05b}.

\section{Summary}
\label{sum}

In this paper we use an analytical model for the dynamics (KA) and the luminosity evolution (KDA) of individual radio galaxies and radio-loud quasars with FRII-type morphologies to explain some of the observed properties of the entire source population. We show that the radio luminosity of the lobes evolves through phases governed by the dominant energy loss mechanism of the radiating, relativistic electrons. While the lobe expands in the quasi-constant density core close to the centre of the host galaxy, the luminosity first rises and then levels out to a constant value as synchrotron losses become dominant. At the same time the spectrum steepens, consistent with the properties of the small CSS sources. As the lobe grows larger than the core radius, the external density decreases and synchrotron losses become less important. Consequently, the spectrum flattens again.

We complement the model for FRII-type objects by presenting a very simplified model for FRI-type lobes. In this approach, the radio luminosity of the FRI-type lobes is constant for most of their lifetime and linearly proportional to the jet power. We also postulate that all radio galaxies and radio-loud quasars start their life with an FRII-type morphology. However, the lobes of objects with weaker jets come into pressure equilibrium with the surrounding gas before the lobe extends beyond the core region. Starting close to the centre of the source, the lobes are then buoyantly replaced by the denser ambient medium which will eventually reach the jet flow itself. Without its protective lobe a weak jet can turbulently disrupt and develop an FRI-type lobe. Somewhat more powerful jets may be able to avoid disruption and to produce another pair of FRII-type lobes. The DDRGs show this kind of structure.

With the combined model for both FR classes we can self-consistently explain the broken power law shape of the RLF. The predicted break luminosity is close to that derived from observed complete samples. The RLF below the break luminosity is assumed to be dominated by FRI-type objects. The slope of the RLF in this region gives the rate at which AGNs producing jets with jet power $Q$ appear in the universe. This birth function is proportional to $Q^{-1.6}$. Above the break luminosity, FRII-type objects dominate and the observed slope of the RLF can be successfully reproduced from the shape of the birth function and the known luminosity evolution.

We also recover the observed distribution of lobe sizes for FRII-type objects. The surprisingly flat size distribution at small sizes is caused by the increasing number of sources developing FRI-type lobes and thereby `dropping out' of the distribution. However, this result depends critically on a significant decrease of the radio luminosity of those sources transforming into FRI-types at the point in time their jets disrupt. 

We explain the dependence of the radio luminosity delineating the FR classes on the optical luminosity of the host galaxy. More luminous hosts are located in denser environments in which more powerful jets can be disrupted. More powerful jets would normally lead to more luminous lobes and so the demarcation line between the FR classes in terms of radio luminosity increases with increasing optical luminosity of the host. 

Finally we discuss the connection between the recently proposed difference in the accretion modes of LERGs and HERGs and their radio morphologies in the context of our model. Our findings are consistent with the idea of radiatively inefficient accretion of hot gas in LERGs and luminous cold gas accretion in HERGs, if radiative efficiency is correlated with jet power. However, differences in the fuelling mechanism between the two classes may pose a problem for our construction of the RLF as we assumed the same birth function for all jet-producing AGN. The problem can be avoided, if the radiative efficiency of the accretion process is regulated by instabilities in the accretion disc itself, independent of the fuelling mechanism. This latter idea is inspired by the observed situation in microquasars. It may provide yet another link between accreting and jet-producing black holes spanning orders of magnitude in black hole mass. 

\section*{Acknowledgements}

The authors wish to thank Martin Hardcastle, Elmar K{\"o}rding, Rob Fender, Julia Riley and Christian Knigge for many helpful discussions. The authors thank the referee, Paul Wiita, for useful and very speedy comments on the original manuscript. CRK thanks PPARC for financial support in the form of a Rolling Grant. PNB would like to thank the Royal Society for generous financial support through its University Research Fellowship scheme.

\def\newblock{\hskip .11em plus .33em minus .07em}

\bibliography{crk}
\bibliographystyle{mn2e}

\appendix

\section{Re-working the KA and KDA models}

In the following we derive expressions for the dynamical evolution and radio synchrotron emission properties of FRII-type radio-loud AGN. Unless stated otherwise, these expressions are identical to the relations given in the papers \citet[][KA]{ka96b} and \citet[][KDA]{kda97a}. Our reason for re-deriving them is to demonstrate how they depend in detail on the most fundamental properties of the jets and their gaseous environments. In particular we concentrate on the jet power, $Q$, the length of the lobe along the jet axis, $D$, the age of the jet flow, $t$, the pressure inside the lobe, $p$, the monochromatic luminosity density, $L_{\nu}$, at a given observing frequency $\nu$ and the `density parameter', $\rho a^{-\beta}$ (see below for an explanation). 

In what follows we necessarily need to frequently reference material in the KA and KDA papers. We abbreviate references in the following way: KA4 refers to equation 4 in KA, while KDA2 refers to equation 2 in KDA.

\subsection{Source dynamics}

In our discussion we concentrate on a single lobe inflated by one of the jets in a radio-loud AGN. The jet is assumed to come into pressure equilibrium with its lobe through a reconfinement shock. After this, the jet flow remains laminar until it enters the hot spot region where its kinetic energy is dissipated. We also assume that the density distribution of the medium external to the lobe is given by a power-law distribution according to
\begin{equation}
\rho_ x = \rho \left( \frac{r}{a} \right) ^{-\beta},
\label{exdens}
\end{equation}
where $\rho$ and $a$ are constants and $r$ measures the distance from the AGN assumed to be located at the centre of the distribution. The power-law form of the density distribution implies that $\rho$ and $a$ are not independent parameters and that the model can only depend on their combination $\rho a^{\beta}$, which we refer to as the density parameter.

With these assumptions, the length of the lobe is given as (KA4 and KA5)
\begin{equation}
D = c_1 \left( \frac{Q}{\rho a^{\beta}} \right)^{1 / \left( 5 - \beta \right)} t^{3/ \left( 5 - \beta \right)}.
\label{length}
\end{equation}
The constant $c_1$ can be written as 
\begin{equation}
c_1 = \left\{ \frac{A^4}{18 \pi} \, \frac{\left( \Gamma _x + 1 \right) \left( \Gamma _l -1 \right) \left( 5 - \beta \right)^3}{9 \left[ \Gamma _l + \left( \Gamma _l - 1 \right) A^2 / 2 \right] - 4 - \beta } \right\}^{1/ \left( 5 - \beta \right) }.
\end{equation}
The ratio of the lobe length and its radius, $A$, stays constant in our self-similar model. Here, we assumed a cylindrical geometry of the lobe. Note that KA uses the aspect ratio of the lobe $R_{\rm T}$ which is denoted $R$ in KDA. Our choice for $A$ implies $A = 2 R_{\rm T} = 2 R$. $\Gamma _l$ and $\Gamma _x$ are the adiabatic indices of the lobe material and the external gas, respectively. Note that this expression for $c_1$ differs slightly from KA25 due to an error in KA first pointed out by \citet{kf98}.

The pressure inside the lobe is given as a function of the source age $t$ by KA20. Using equation (\ref{length}) and following the discussion on the expansion of the lobe leading up to KA38, we find
\begin{equation}
p = f_p \left( \rho a^{\beta} \right) ^{1/3} Q^{2/3} D^{\left( - 4 - \beta \right) /3},
\label{pressure}
\end{equation}
with 
\begin{equation}
f_p = \frac{18 c_1^{2 \left( 5 - \beta \right) / 3} }{\left( \Gamma _x +1 \right) \left( 5 - \beta \right)^2 A^2}.
\end{equation}
Note here that by substituting equation (\ref{length}) into equation (\ref{pressure}), we can express the pressure in the lobe as a function of the source age,
\begin{equation}
p \propto \left( \rho a^{\beta} \right)^{3 / \left( 5 - \beta \right)} Q^{\left( 2 - \beta \right) / \left(5 - \beta \right)} t^{\left(-4 - \beta \right) / \left( 5 - \beta \right)}.
\end{equation}
This shows that if $\beta \sim 2$, then at a given point in time the pressure inside the lobe is the same in all sources located in environments of similar density, regardless of the jet power.

\subsection{Synchrotron emission}

We now turn to the processes giving rise to the radio synchrotron emission of the lobe. We have to take into account the energy losses of the relativistic electrons inside the lobe. These depend on the time the electrons have spent inside the lobe and so depend not only on the age of the jet flow, $t$, but also on the time $t_{\rm i} < t$ at which they were injected into the lobe. The approach of KDA is to split the lobe into small volume elements ${\rm d} V$ which are `labelled' with their injection time $t_{\rm i}$. The evolution of the electron population inside these ${\rm d} V$ can then be traced individually in a self-consistent manner. KDA show that for adiabatic expansion, ${\rm d} V \propto t^{a_1}$ with (see KDA14)
\begin{equation}
a_1 =  \frac{4+\beta}{\Gamma _l \left( 5 - \beta \right)}.
\end{equation}

The radiating relativistic electrons are accelerated at the end of the jet flow such that their total energy and the energy stored in the magnetic field at least initially follow the requirements of the minimum energy condition \citep[e.g.][]{ml94}. The energy distribution of the relativistic electrons follows a power-law at time $t_{\rm i}$, when they are injected into the lobe,
\begin{equation}
n \left( \gamma \right) \,{\rm d} \gamma = n_0 \left( t_{\rm i} \right) \gamma ^{-m} \, {\rm d} \gamma,
\label{endist}
\end{equation}
with $\gamma$ the Lorentz factor of the electrons. At the time $t_{\rm i}$ the energy densities of the magnetic field, $u_{\rm B}$, and that of the electrons, $u_{\rm e}$, is then fixed to (KDA15)
\begin{eqnarray}
u_{\rm B} \left( t_{\rm i} \right) & = & \frac{p \left( t_{\rm i} \right)}{\left( \Gamma _l - 1 \right) \left( k + 1 \right) \left( r + 1 \right)}, \label{magdens}\\
u_{\rm e} \left( t _{\rm i} \right) & = & \frac{r p \left( t _ {\rm i } \right)}{\left( \Gamma _l - 1 \right) \left( r + 1 \right)},
\end{eqnarray}
where $r = \left( m + 1 \right) / 4$ and $k$ is the ratio of the energy stored in non-radiating particles and the sum of the energy in the magnetic field and the relativistic electrons. Since $p (t) = p(t_{\rm i}) \left( t / t_{\rm i} \right)^{-\Gamma _l a_1}$ and $u_{\rm B} (t) = u_{\rm B} ( t_{\rm i}) \left( t / t_{\rm i} \right)^{-4 a_1 / 3}$, the energy density of the magnetic field in the lobe at time $t$ is
\begin{equation}
u_{\rm B} (t) = \frac{p(t)}{\left( \Gamma _l -1 \right) \left( k + 1 \right) \left( r + 1 \right)} \left( \frac{t}{t_{\rm i}} \right)^{a_1 \left( \Gamma _l - 4/3 \right)}.
\label{ub}
\end{equation}
The energy density and hence the strength of the magnetic field is not uniform throughout the lobe, if $\Gamma _l \ne 4/3$. Note here that we do not consider the case where the magnetic pressure in the lobe follows a non-relativistic equation of state (case C in KDA). Such a situation would require a very special behaviour of the MHD turbulence inside the lobe constantly removing energy from the magnetic field during the expansion. Hence we set $\Gamma _{\rm B} = 4/3$ wherever the adiabatic index of the magnetic field appears in KDA.

The Lorentz factor of a relativistic electron emitting radiation predominantly at frequency $\nu$ is in SI units
\begin{equation}
\gamma = \sqrt{ \frac{2 \pi m_{\rm e} \nu}{e B}},
\label{lorentz}
\end{equation}
where the strength of the magnetic field is related to the energy density of the field by
\begin{equation}
B = \sqrt{ 2 \mu_0 u_{\rm B}},
\end{equation}
with $\mu_0$ the permeability of the vacuum. From the discussion of the evolution of $u_{\rm B}$ it follows that
\begin{equation}
\gamma = f_{\gamma} p ^{-1/4} \left( \frac{t}{t_{\rm i}} \right)^{a_1 \left( 4/3 - \Gamma _l \right)/4},
\label{gammat}
\end{equation}
where the constant $f_{\gamma}$ can be found from equations (\ref{magdens}) and (\ref{lorentz}) as
\begin{equation}
f_{\gamma} = \sqrt{ \frac{2 \pi m_{\rm e} \nu}{e} \sqrt{\frac{\left( \Gamma _l -1\right) \left( k +1 \right) \left( r + 1 \right)}{2 \mu_0 f_p}}}.
\end{equation}

The normalisation of the energy distribution of the electrons, equation (\ref{endist}), is set by the energy density $u_{\rm e} (t_{\rm i})$ (see KDA8) and can be written as
\begin{equation}
n_0 \left( t_{\rm i} \right) = f_n p \left( \frac{t}{t_{\rm i}} \right)^{a_1 \Gamma_l},
\end{equation}
where the constant $f_n$ is related to the maximum and minimum Lorentz factors $\gamma _{\rm max}$ and $\gamma _{\rm min}$ at the time of injection by
\begin{equation}
f_n = \frac{r}{\left(\Gamma _l -1\right) \left( r + 1 \right) m_{\rm e} c^2} \left( \frac{\gamma _{\rm min}^{2-m} - \gamma _{\rm max}^{2-m}}{m-2} - \frac{\gamma _{\rm min}^{1-m} - \gamma _{\rm max}^{1-m}}{m-1} \right)^{-1},
\end{equation}
for $m \ne 2$ and
\begin{equation}
f_n = \frac{r}{\left(\Gamma _l -1\right) \left( r + 1 \right) m_{\rm e} c^2} \left( \ln \frac{ \gamma _{\rm max}}{\gamma _{\rm min}} - \frac{1}{\gamma_{\rm min}} + \frac{1}{\gamma_{\rm max}} \right)^{-1},
\end{equation}
for $m=2$.

We now use the results above to express the monochromatic radio synchrotron luminosity in such a way to emphasise its dependence on the most important source parameters. For this we substitute the expressions derived above into equation KDA16. While KDA16 gives the radio emission per unit solid angle, $P_{\nu}$, we assume that the emission is isotropic and so the radio luminosity derived here is given by $L_{\nu} = 4 \pi P_{\nu}$.The result is
\begin{equation}
L_{\nu} = f_L Q p^{\left( m + 1 \right) / 4} t  \int_{x_{\rm min}}^1 x^{\epsilon} \left[ 1 - \gamma g \left( x, t \right) \right]^{m-2} \, {\rm d} x,
\label{lumin}
\end{equation}
where 
\begin{equation}
f_L = \frac{2 A^{2 \left( 1-\Gamma _l  \right) / \Gamma_l} f_{\gamma}^{3-m} f_n}{3 \left( k +1 \right) \left( r + 1 \right)} \frac{\sigma_{\rm T} c}{\nu},
\end{equation}
with $\sigma _{\rm T}$ the Thomson cross-section, and
\begin{equation}
\epsilon = \left( 1 + \frac{m}{3} -\Gamma_l \right) a_1+\frac{a_1}{4} \left( \Gamma_l -\frac{4}{3} \right) \left( 3 - m \right).
\label{eps}
\end{equation} 
We have also changed the integration variable from $t_{\rm i}$ in KDA16 to $x=t_{\rm i} / t$. The part of the lobe injected at the time $t_{\rm i}$ corresponding to $x_{\rm min}$ is the oldest part of the lobe still contributing to the overall emission at the current time $t$. The value of $x_{\rm min}$ is determined by the energy losses of the relativistic electrons and these are represented by the function $g \left( x , t \right)$. We will discuss this function and the determination of $x_{\rm min}$ next. However, there is no general analytic solution for the integral in equation (\ref{lumin}).

\subsection{Energy losses of the relativistic electrons}
\label{loss}

Inside the lobe, the relativistic electrons are subject to energy losses due to the expansion of the lobe, the emission of synchrotron radiation and the inverse Compton scattering of CMB photons. Their Lorentz factor evolves according to
\begin{equation}
\frac{{\rm d} \gamma}{{\rm d} t} = - \frac{a_1}{3} \frac{\gamma}{t} - \frac{4 \sigma_{\rm T}}{3m_{\rm e} c} \gamma^2 \left( u_{\rm B} + u _{\rm CMB} \right),
\label{dotgamma}
\end{equation}
where $u_{\rm CMB}$ is the energy density of the CMB photons at the source redshift. We use the initial condition that the Lorentz factor of an electron is $\gamma_{\rm i}$ at the time of injection into the lobe $t_{\rm i}$. At time $t$ the Lorentz factor of the same electron is $\gamma$. The solution of this differential equation is then given by (compare with KDA6 and KDA7)
\begin{equation}
\gamma = \frac{x^{a_1/3} \gamma _{\rm i}}{1+ x^{a_1/3} \gamma_{\rm i} g \left( x, t \right) },
\label{gamma}
\end{equation}
where
\begin{equation}
g \left( x , t \right) = \frac{4 \sigma _{\rm T}}{3 m_{\rm e} c} t \left[ \frac{u_{\rm B} \left(t \right)}{a_3} \left( 1 - x^{a_3} \right) + \frac{u_{\rm CMB}}{a_4} \left( 1 - x^{a_4} \right) \right],
\end{equation}
for $a_1 \ne  3/ 5$, with $a_3 = 1 - 5 a_1 / 3$ and $a_4 = 1 - a_1 /3$, and
\begin{equation}
g \left( x , t \right) = \frac{4 \sigma _{\rm T}}{3 m_{\rm e} c} t \left[ - u_{\rm B} \left( t \right) \ln x + \frac{5}{4} u_{\rm CMB} \left( 1 - x^{4/5} \right) \right],
\label{a135}
\end{equation}
for $a_1 = 3/5$. The latter case occurs when $\Gamma_l = 4/3$ and $\beta = 0$ or $\Gamma_l = 5/3$ and $\beta = 1/2$. The function $g \left( x, t \right)$ is equal to $t^{a_1/3} a_2 \left( t, t_{\rm i} \right)$, with $a_2$ the function defined in KDA7.

At time $t$ the Lorentz factor of the electrons emitting radiation at the observing frequency $\nu$ is given by equation (\ref{lorentz}). For $\Gamma_l \ne 4/3$, this Lorentz factor $\gamma _{\nu}$ depends on the injection time $t_{\rm i}$ and is therefore in general not uniform throughout the lobe. The energy of all electrons is decreasing monotonically. Therefore the oldest part of the lobe still contributing to the emission at frequency $\nu$ is defined by the condition that electrons with a Lorentz factor $\gamma _{\nu}$ at the current time $t$ had a Lorentz factor $\gamma _{\rm i} = \gamma _{\rm max}$ at the time they were injected into the lobe, $t_{\rm i}$. Thus the lower limit of the integration in equation (\ref{lumin}), $x_{\rm min}$, is defined by
\begin{equation}
\gamma _{\nu} = f _{\gamma} p ^{-1/4} x_{\rm min}^{a_1 \left( \Gamma _l - 4/3 \right) / 4} = \frac{x_{\rm min}^{a_1/3} \gamma _{\rm max}}{1+ x_{\rm min}^{a_1/3} \gamma_{\rm max} g \left( x_{\rm min}, t \right) }.
\label{xmin}
\end{equation}
For very large $\gamma_{\rm max}$ this can be simplified to
\begin{equation}
f_{\gamma} p ^{-1/4} \sim \frac{x_{\rm min}^{a_1 \left( 4/3 - \Gamma _l \right) / 4}}{g \left( x_{\rm min} , t \right)},
\end{equation}
but even in this case the relation can usually not be written as an explicit expression for $x_{\rm min}$.

\end{document}